\documentclass[prb,floatfix,twocolumn,showpacs,amsfonts]{revtex4}
\usepackage{graphicx}
\usepackage{bm}
\usepackage{amsmath2000}
\begin{document}
\title{Collective modes in the d-density wave state of the cuprates}
\author{Sumanta Tewari and Sudip Chakravarty}
\email[First author: ]{sumanta@physics.ucla.edu}
\email[second author: ]{sudip@physics.ucla.edu}
\affiliation{Department of Physics, University of California Los Angeles, Los Angeles, CA
90095--1547}
\date{\today}
\begin{abstract}
Using a functional integral formulation, we analyze the collective modes
in the $d$-density wave state. Since only discrete
symmetries are broken, no massless phase mode is present. The only relevant
fluctuation is the amplitude fluctuation of the order parameter, which
is treated at the gaussian level above the long range order. These
fluctuations give rise to a massive mode, the analog of the $U(1)$ Higgs, or amplitude  mode
at wave-vector $(\pi,\pi)$. Neutrons in inelastic neutron scattering experiments 
couple to this mode via the orbital currents and are likely to produce
direct measurements of this amplitude boson.
\end{abstract}
\maketitle

 Much attention has recently been paid to a discrete symmetry broken state,
 the $d$-density wave (DDW), as a possible candidate for the 
 anomalous normal state of the underdoped cuprate superconductors. \cite{CLMN,CK}
 Below the pseudogap temperature scale, $T^{*}$, the system is conjectured
 to condense into the DDW state, which is  responsible for much of the 
 observed physics of the pseudogap phase. Further down in temperature,
 below the superconducting transition temperature $T_{c}$, the DDW and the
 $d$-wave superconductor (DSC) are expected to coexist and compete for
 the same regions on the Fermi surface. In a range of the hole-doping
 concentration, $x$, suitable for making up the underdoped systems, the competition
 between the two orders consistently explains the unique variation of the superconducting
 order parameter with doping. \cite{CLMN}
 
 The DDW state, under various names and guises, appeared in the literature
 on many occasions.\cite{HR,Schulz,N1,N2,Kotliar,MA,Chetan,Patrick} More elaborate treatments of some 
 of the  properties of the system, most relevant to 
  experiments on the high-$T_{c}$ superconductors, were discussed  recently.
\cite{Hsu,Sumanta1,Sumanta2,Sudip}
 However, these authors, for the most part, treated the system in the framework
 of Hartree-Fock mean field theory, one consequence of which was that, all information
 about the collective modes - and experimental implications thereof - remained uninvestigated.
 In this paper, we intend to fill this gap by elucidating the structure of the collective modes
 and pointing out how they can actually be
 detected in experiments.

The DDW state has plaquette-currents, alternating 
clockwise-anticlockwise in the neighboring plaquettes of an underlying
square lattice. Thus, it gives rise to tiny orbital moments. The relevant order-parameter characterizing the
$d$-density wave is a paricle-hole spin-singlet pair, condensing at an wavevector ${\bf Q}=(\pi,\pi)$.
The bipartite square lattice band-structure
is equivalent under the transformation ${\bf Q}\rightarrow -{\bf Q}$ and this
forces the order parameter to be imaginary, \cite{Chetan} a fact which is
ultimately responsible for the breaking of the time-reversal symmetry and the resulting
spontaneous bond-currents. The phase of the order parameter being fixed
by the specific symmetries it breaks, the only natural collective excitation
one expects is the amplitude fluctuation of the order-parameter, which
will give rise to a finite frequency mode, the amplitude mode, centered at ${\bf Q}$.
We will explicitly calculate the frequency to be $2\Delta_{{\rm DDW}}$, where $\Delta_{{\rm DDW}}$ is
the DDW single particle gap at $(\pi,0)$, which can be estimated from photo-emission
experiments above $T_{c}$.
We will also show how inelastic neutron scattering, coupling to the
bond-currents, should directly reveal this mode. In fact, well-defined, but damped
features at  wave-vector ${\bf Q}$ and energy  comparable to $2\Delta_{{\rm DDW}}$
have indeed been seen above $T_{c}$ in inelastic neutron scattering of the cuprates.\cite{FongMook}
These damped peaks are in sharp  contrast to the  
resonant peak below $T_{c}$, which
occurs at the same wave-vector and comparable energy, but is resolution-limited in 
frequency. This opens up a tantalizing
possibility that the amplitude mode of the DDW state, which is the analog of the 
Higgs mode of $U(1)$ gauge theory, may actually have already been
observed in experiments. Clearly, the issue needs further investigation.

In order to analyze the collective modes, let us
 start with the  Hamiltonian in momentum space,
\begin{equation} 
H = \sum_{{\bf k}, \sigma} (\epsilon_{{\bf k}}-\mu) c_{{\bf k} \sigma}^{\dagger} c_{{\bf k} \sigma} -
    \sum_{{\bf k},{\bf k^{\prime}}, {\bf q},\alpha,\beta} g_{{\bf k}{\bf k}^\prime}^{{\bf q}} 
       c_{{\bf k}+{\bf q} \alpha}^{\dagger} 
       c_{{\bf k} \alpha} 
       c_{{\bf k}^\prime \beta}^{\dagger}
       c_{{\bf k}^\prime+{\bf q} \beta},
\label{Hamiltonian1}      
\end{equation}
where $c_{{\bf k} \sigma}^{\dagger}$ creates a particle with momentum ${\bf k}$ and spin $\sigma$,
$\epsilon_{{\bf k}}$ is the free electron band structure $-2t(\cos k_{x}+\cos k_{y})$, $\mu$ is the
chemical potential, 
and $g_{{\bf k}{\bf k}^\prime}^{{\bf q}}$
is the interaction matrix element. A more realistic
band structure would include
the effect of a next-neighbor hopping $t'$.
A small $t'$ does not change our results qualitatively, so we ignore it for simplicity.
In the half filled limit, $\mu=0$; it  takes
non-zero negative values when holes are doped into the system.    
In this paper we will work in units where the lattice
constant is unity and will also set
$\hbar = c = k_{B} = 1$.

 The particle-hole spin-singlet $d$-density wave order parameter is
 anisotropic in $k$-space, \begin{equation}\langle c_{{\bf k},\alpha}^{\dagger}c_{{\bf k}+{\bf
Q},\beta}\rangle=
 i\frac{\Phi_{{\bf Q}}}{2}f_{{\bf k}}\delta_{\alpha \beta},\end{equation} where 
 $\Phi_{{\bf Q}}$ is the magnitude of the order parameter, and 
 $f_{{\bf k}}=(\cos k_{x}-\cos k_{y})$. So, we will take the
 effective part of the full interaction factorizable as \begin{equation} g_{{\bf k}{\bf k}^\prime}^{{\bf q}}=
 f_{{\bf k}} f_{{\bf k}^\prime}g.\end{equation} This would enable us to always maintain 
 the $d_{x^2-y^2}$-wave symmetry of the order parameter and discuss long wavelength 
 fluctuations around it. For a discussion
 of the DDW state, we will 
 ignore the other parts of the 
 interaction and the fluctuations of other symmetries. 

    The Hamiltonian now takes the form
    \begin{equation} 
H = \sum_{{\bf k}, \sigma} (\epsilon_{{\bf k}}-\mu) c_{{\bf k} \sigma}^{\dagger} 
c_{{\bf k} \sigma} - g^{-1}
    \sum_{{\bf q}} \widehat{\Delta}^{\dagger}({\bf q}) \widehat{\Delta}({\bf q}),
    \label{Hamiltonian2}
\end{equation} 
where,
\begin{equation} 
\widehat{\Delta}({\bf q})=g\sum_{{\bf k},\alpha}f_{{\bf k}}c_{{\bf k} \alpha}^{\dagger}
c_{{\bf k}+{\bf q} \alpha}.
\label{Delta}
\end{equation}
Note, for ${\bf q}={\bf Q}$, a commensurate wave-vector, 
$\widehat{\Delta}({\bf Q})$ is antihermitian,\cite{Chetan} and is, in operator form,
the DDW gap parameter.

In functional integral form, the partition function is
 \begin{equation} \label{Partition1}
Z = {\cal N} \int {\cal D} c^+ {\cal D} c e^{-S_0} ,
\end{equation}
where
\begin{widetext}
\begin{equation} 
S_0 \equiv \int_0^{\beta} d \tau
    \left( \sum_{{\bf k} \sigma} ( \partial_\tau + \epsilon_{{\bf k}}-\mu ) c_{{\bf k} \sigma}
    ^{\dagger} c_{{\bf k} \sigma}
      - g^{-1} \sum_{{\bf q}} \widehat{\Delta}^{\dagger}({\bf q}) \widehat{\Delta}({\bf q}) \right),
 \label{Action1}     
\end{equation}
\end{widetext}
and ${\cal N}$ is a normalization constant.
By introducing an auxilliary boson field $\Delta_{{\bf q}}$ through the Hubbard-Stratonovich
transformation, the interaction term is made quadratic in the Fermion fields.
In frequency space, the new action is given by
\begin{equation} 
S_{0}^{\prime} \equiv \sum_{{\bf k},\omega} \frac{\beta}{g} \Delta^{\dagger}({\bf k},\omega) \Delta({\bf k},\omega)
 +\sum_{{\bf k}_{1}{\bf k}_{2} \omega_{1}\omega_{2}}\Psi_{{\bf k}_{1} \omega_{1}}^{\dagger}
	 M_{\omega_{1},\omega_{2}}^{{\bf k}_{1},{\bf k}_{2}}\Psi_{{\bf k}_{2} \omega_{2}},
\label{Action2}	 
\end{equation}	 
where $\Psi_{{\bf k} \omega}^{\dagger}$ is a two-component vector, $\Psi_{{\bf k} \omega}^{\dagger}=
( c_{{\bf k} \omega}^{\dagger}, c_{{\bf k}+{\bf Q},\omega }^{\dagger} ) $, and $M$ is a 2$\times$2 matrix given by,
\begin{widetext}
\begin{equation}
M_{\omega_1,\omega_2}^{{\bf k}_1,{\bf k}_2}  \equiv \beta \delta_
{{\bf k}_1,{\bf k}_2} \delta_{\omega_1,\omega_2}[(-i\omega_2 -\mu)I + \epsilon_{{\bf k}_2}\sigma_3]
-\frac{\beta}{2}[f_{{\bf k}_{1}} \Delta_{{\bf k}_2+{\bf Q}-{\bf k}_1,\omega_2-\omega_1}
^{\dagger}\sigma^{+}+ f_{{\bf k}_2} \Delta_{{\bf k}_1+{\bf Q}-{\bf k}_2,\omega_1-\omega_2}\sigma^{-}],
\end{equation}
\end{widetext}
where the $\sigma$'s are the Pauli matrices acting in the space of $\Psi_{{\bf k} \omega}^{\dagger}$, and,
$\sigma^{\pm}=\sigma_1 \pm i\sigma_2$.


In Eq.~\ref{Action2}, the full Brillouin zone has been folded to the magnetic,
or reduced Brillouin zone (RBZ), to facilitate the discussion of the DDW state;
 the momentum integrations are restricted to the RBZ,
 and a spin-sum over the Fermionic variables is implied.
After integrating out the Fermion fields, the partition function is given by,
\begin{equation} \label{Partition2}
Z = {\cal N^{\prime}} \int {\prod_{{\bf k},\omega}} d\Delta^{\dagger}_{{\bf k},\omega} d\Delta_{{\bf k},\omega} 
    e^{-S} ,
\end{equation}
where, 
\begin{equation} \label{Action3}
S \equiv \sum_{{\bf k},\omega} \frac{\beta}{g} \Delta^{\dagger}({\bf k},\omega) \Delta({\bf k},\omega)
            - 2\ln \det M,
\end{equation} 
and ${\cal N}^{\prime}$ is now a different normalization constant.
The factor of two in the second term of $S$ comes from the spin-summation.
By taking $\partial S/\partial \Delta^{\dagger}({\bf k},\omega) = 0$ and $\partial S/
\partial \Delta({\bf k},\omega) = 0$, we find the saddle point,
\begin{equation} \label{saddlepoint1}
\Delta({\bf k},\omega) = 0,\;\text{for} \;{\bf k} \neq {\bf Q}, \;\text{or}\; \omega \neq 0,
\end{equation}
and,
\begin{equation} \label{saddlepoint2}
\Delta({\bf Q},0) = \Delta_{{\bf Q}} = g\sum_{{\bf k}} f_{{\bf k}}^2  \left(\frac{\Delta_{{\bf Q}}}{E_{{\bf
k}}}\right)\Theta(\mu+E_{{\bf k}}).
\end{equation}
Here, $E_{{\bf k}} \equiv \sqrt{\epsilon_{{\bf k}}^2 + f_{{\bf k}}^2|\Delta_{{\bf Q}}| ^2}$ and $\Theta(x)$ is a unit step
function. Equation~\ref{saddlepoint2} is the $d$-density wave mean-field gap equation. Notice, with this definition
of $E_{{\bf k}}$, the DDW single-particle gap at $(\pi,0)$, 
$ \Delta_{{\rm DDW}}=2|\Delta_{{\bf Q}}|$.

As $\widehat{\Delta}({\bf Q},0)$ is antihermitian, the gap parameter
$\Delta({\bf Q},0)$ must be pure imaginary. Hence, though $\Delta({\bf Q},0) = \Delta_{{\bf Q}} e^{i\theta}$
satisfies the gap equation equally well, the phase of the order parameter is locked to
$\pm \frac{\pi}{2}$.\cite{Chetan} Thus, there are
no massless phase modes in the long-wavelength collective fluctuations
of the order parameter. Put another way, since the order parameter breaks only  
discrete symmetries, there are no Goldstone modes in the system. For an incommensurate
version of the DDW state, where the ordering wave-vector is not commensurate with
the lattice, there is a collective massless mode, the sliding density wave mode. But
this mode will be pinned by the impurities as well, in second order of the impurity potential.
    
    For an analysis of the amplitude mode, the task at hand is to expand the action functional $S$
around the $d$-density wave saddle point. By doing so, we will take account
of fluctuations of the order parameter amplitude in the  plaquettes, implying
fluctuations in the bond-currents. This will impose fluctuating non-zero values for $\Delta({\bf k},\omega)$,
at wave-vectors ${\bf k} \neq {\bf Q}$, and frequencies $ \omega \neq 0$.
Expanding $S$ around the saddle point, to the second order in $\Delta_{{\bf k}+{\bf Q},\omega}$,
\begin{widetext}
\begin{equation} \label{Perturb action1}
\delta S  =   \sum_{{\bf k},\omega}^{\prime}( A_{{\bf k}+{\bf Q},\omega}\Delta^{\dagger}_
{{\bf k}+{\bf Q},\omega}\Delta_{{\bf k}+{\bf Q},\omega}+A_{-{\bf k}-{\bf Q},-\omega}
\Delta^{\dagger}_{-{\bf k}+{\bf Q},-\omega}\Delta_{-{\bf k}+{\bf Q},-\omega}
   +  B_{{\bf k}+{\bf Q},\omega} (\Delta^{\dagger}_{{\bf k}+{\bf Q},\omega}\Delta^{\dagger}_{-{\bf k}+{\bf Q},-\omega}
   +\Delta_{-{\bf k}+{\bf Q},-\omega}\Delta_{{\bf k}+{\bf Q},\omega} ) ), 
\end{equation}
\end{widetext}
where the prime over the summation indicates exclusion of the points 
$({\bf k} =0)$ and $(\omega =0)$. $A$ and $B$ are given by,
\begin{widetext}
\begin{eqnarray} \label{A1}
A_{{\bf k}+{\bf Q},\omega}& \equiv & \frac{\beta}{2g}+\sum_{{\bf k}',\omega'}f_{{\bf k}'}^2
\frac{(i\omega'-\mu-\epsilon_{{\bf k}'})[i(\omega'+\omega)-\mu+
\epsilon_{{\bf k}'+{\bf k}}]}{[(i\omega'+\mu)^2-E_{{\bf k}'}^2][(i(\omega'+\omega)+\mu)^2-E_{{\bf k}'+
{\bf k}}^2]}\\
B_{{\bf k}+{\bf Q},\omega} &\equiv& \sum_{{\bf k}',\omega'} \frac{f_{{\bf k}'}^2 f_{{\bf k}'+
{\bf k}}^2 \Delta_{{\bf Q}}^2}
{[(i\omega'+\mu)^2-E_{{\bf k}'}^2][(i(\omega'+\omega)+\mu)^2-E_{{\bf k}'+
{\bf k}}^2]}
\end{eqnarray} 
\end{widetext}

Noting that $\Delta_{{\bf k}+{\bf Q},\omega}$ is the Fourier component of a pure imaginary field, we 
have the property,
\begin {equation} 
\Delta^{\dagger}_{-{\bf k}+{\bf Q},-\omega}=-\Delta_{{\bf k}+{\bf Q},\omega}.
\label{prop}
\end {equation}
Using this, and also the fact that $\Delta_{{\bf Q}}^2=-|\Delta_{{\bf Q}}|^2$ 
in the definition of $B_{{\bf k}+{\bf Q},\omega}$, we collapse 
Eq.~\ref{Perturb action1}
onto, 
\begin{eqnarray} \label{Perturb action2}
\delta S  &=&   \sum_{{\bf k},\omega}^{\prime}( A_{{\bf k}+{\bf Q},\omega}+A_{-{\bf k}-{\bf Q},-\omega}+
2\widetilde{B}_{{\bf k}+{\bf Q},\omega})\nonumber\\
&\times&\Delta^{\dagger}_{{\bf k}+{\bf Q},\omega}\Delta_{{\bf k}+{\bf Q},\omega},
\end{eqnarray}
where $\widetilde{B}_{{\bf k}+{\bf Q},\omega}$ is $B_{{\bf k}+{\bf Q},\omega}$ with $\Delta_{{\bf Q}}$
replaced by $|\Delta_{{\bf Q}}|$ in the numerator.

The dispersion for the collective mode 
can be read off by equating the fluctuaion kernel to zero,\cite{popov}
\begin{equation}
\label{amplitude1}
 A_{{\bf k}+{\bf Q},\omega}+A_{-{\bf k}-{\bf Q},-\omega}+2\widetilde{B}_{{\bf k}+{\bf Q},\omega}=0.
\end{equation}

In the case of a superconductor,
there is no symmetry property, such as Eq.~\ref{prop}, for the fluctuation fields. Gaussian fluctuations
around the superconducting saddle point has a form very similar to Eq.~\ref{Perturb action1},\cite{Lan}
and the collective modes are given by the determinantal equation, 
$A^{\prime}_{{\bf k},\omega}A^{\prime}_{-{\bf k},-\omega}-B^{\prime}_{{\bf k},\omega}
 B^{\prime}_{-{\bf k},-\omega} = 0$. Using the properties, $A^{\prime}_{{\bf k},\omega}=A^{\prime}_{-{\bf k},-\omega}$ 
 and $B^{\prime}_{{\bf k},\omega}=B^{\prime}_{-{\bf k},-\omega}$,\cite{Lan} which are specific to a superconductor, this
 splitts up into two equations, $A^\prime({\bf k},\omega)-B^\prime({\bf k},\omega)=0$, 
 and $A^\prime({\bf k},\omega)+B^\prime({\bf k},\omega)=0$. These are precisely the two equations
 giving the phase and the amplitude mode dispersions, respectively. The phase mode is 
 massless, and the massive amplitude mode is centered at ${\bf q}=(0,0)$. In the case of DDW, because of the 
 symmetry property of the fluctuation fields which leaves the order parameter 
 purely imaginary, one of the equations drops out,
 and, we are left with just one equation, that for the amplitude mode.
 
 Coming back to Eq.~\ref{amplitude1}, in the case of DDW,
writing
\begin{equation} 
\frac{\beta}{g}=-2\sum_{ {\bf k}',\omega'} \frac{f_{{\bf k}'}^2}
{(i\omega'+\mu)^2-E_{{\bf k}'}^2}
\end{equation}
following the gap equation, in the limit ${\bf k}=0$, the frequency 
$\omega$ for the amplitude mode is given by,  
\begin{equation}
\sum_{ {\bf k}',\omega'} f_{{\bf k}'}^2 \frac{\omega^2+4 f_{{\bf k}'}^2 
|\Delta_{\bf Q}|^2-8i\omega'\mu}{[(i\omega'+\mu)^2-E_{{\bf k}'}^2][(i(\omega'+\omega)+\mu)^2-E_{{\bf k}'
}^2]} =0.
\label{imp1}
\end{equation}

After the internal frequency summation, the last equation, at zero temperature, reduces to,
\begin{equation}
\sum_{ {\bf k}'}\frac {f_{{\bf k}'}^2}{E_{{\bf k}'}}\Theta(\mu+E_{{\bf k'}}) \frac{\omega^2+4 f_{{\bf k}'}^2 
|\Delta_{\bf Q}|^2+8\mu^{2}}{(\omega^2+4E_{{\bf k}'}^2)} =0.
\label{imp11}
\end{equation}

Let us first analyze the mode-frequencies for the simpler case of an $s$-wave density wave,
the charge density wave (CDW). In this case, $f_{{\bf k}}=1$, and $ \Delta_{{\bf Q}}=\Delta$,
a real quantity. Even if the order parameter is real, the last equation, with the substitution
of $f_{{\bf k}}$ by $1$, still gives the 
frequencies of the amplitude mode. This is because, for a real order parameter,
Eq.~\ref{prop} holds with a $+$ sign, and, $\widetilde{B}_{{\bf k}+{\bf Q},\omega}$
and $B_{{\bf k}+{\bf Q},\omega}$ are same. Consequently, one can still collapse Eq.~\ref{Perturb action1}
on to Eq.~\ref{Perturb action2}, and get the same equation for the amplitude mode. 
Considering, then, Eq.~\ref{imp11} for CDW, we note that, for $\mu=0$, the roots are at $\omega=\pm i2\Delta$.
This implies the mode-frequencies at $E=i\omega=\pm2\Delta$. The modes come in pair, for the absorptive
and emissive responses, respectively. For increasing values of $|\mu|$, the modes are
shifted, finally, for large values of $|\mu|$, the frequencies scale with $\pm 2\sqrt{2}|\mu|$.

Returning to the case of DDW, where $f_{{\bf k}}$ is nontrivial, we have to perform
the internal momentum integral numerically. First, after some simple manipulations,
Eq.~\ref{imp11} is brought to the dimensinless form,
\begin{equation}
I(z)\equiv\sum_{ {\bf k}'}\frac {f_{{\bf k}'}^2}
{[e_{{\bf k}'}^{2}+f_{{\bf k}'}^{2}]^{\frac{1}{2}}}\Theta(\mu+E_{{\bf k'}}) \frac{z^{2}+ f_{{\bf k}'}^2 
+2(\frac{\mu}{|\Delta_{{\bf Q}}|})^{2}}{z^2+e_{{\bf k}'}^{2}+f_{{\bf k}'}^{2}} =0,
\label{imp111}
\end{equation}
where, $z=\frac{\omega}{2|\Delta_{{\bf Q}}|}$, and $e_{{\bf k}'}=\frac{\epsilon_{{\bf k}'}}{|\Delta_{{\bf Q}}|}$.

In analogy with the $s$-wave case,
one can easily see from here that, at large $\mu/|\Delta_{{\bf Q}}|$, where the
$f_{{\bf k}}^{2}$ term in the numerator can be neglected in comparison, the roots of the last
equation are at $z\simeq \pm i \sqrt{2}\frac{\mu}{|\Delta_{{\bf Q}}|}$, i.e., at real
frequencies $E=i\omega\simeq \pm 2 \sqrt{2}\mu$. If $\mu$ behaves with doping of
holes in the system in a manner similar to a Fermi liquid, then it can be quite large,
and eventually scales with $t$, the largest scale in the problem. In such a case, the amplitude mode in
the DDW system is at very high energy and is virtually inaccesible in thermal
neutron scattering. Also, a mode so high up in energy is expected to be
heavily damped, since there is a large phase-space
of excitations it can decay into. Consequently, one can not 
in the case of large $|\mu|$ 
meaningfully talk about the amplitude mode in connection with experiments.
But a large value of the chemical potential may not be the case
in the underdoped regime of the cuprates,\cite{Sumanta1} where
various kinds of charge inhomogeneities and impurities may actually pin the
doped holes. Indeed, there are some
indications\cite{Ino} that $\mu$ remains pinned close to zero in these systems.
Under such a scenario, the shift of the mode-frequencies due to chemical potential
is small, and the results are qualitatively similar to the case of half-filling,
$\mu=0$. Thus, in order to make contact with experiments, we will only discuss the
case of chemical potential pinned to zero, assuming it does not shift
appreciably with doping in the underdoped regime of the cuprates.

Assuming $\mu=0$, and replacing $z$ by $iu$ in the left hand side of Eq.~\ref{imp111}, 
we have numerically performed the momentum integral over the RBZ. The reason for replacing $z$ by 
$iu$ is that, we will eventually look for roots of the equation on the imaginary axis 
such that the mode-frequencies
after analytic continuation become real.  
In figure 1, we show the results of the integration plotted as a function of $u$.
\begin{figure}[bht] 
\includegraphics[scale=0.5]{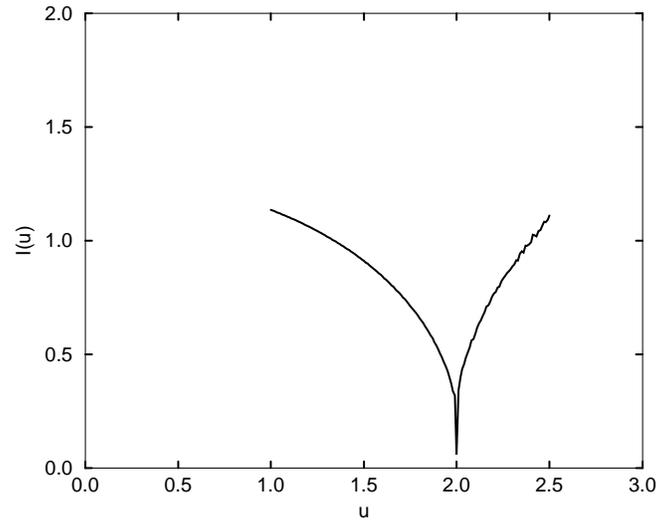}   
\caption 
{Results of the integral given in Eq.~\ref{imp111}, $I(u)$, plotted
as a function of $u$, where $u=-iz$. The parameters used
are $t=0.25$ eV, $|\Delta_{{\bf Q}}|=0.01$ eV, and $\mu=0$. $I(u)$ has a root at $u=2$. 
Since it is an even function of $u$, the other root, not shown here, occurs at 
$u=-2$.} 
\label{fig:1} 
\end{figure}

 Calling the dimensionless  
function of $u$ in Eq.~\ref{imp111} $I(u)$, we note that it has a root at $u=2$. Since $I(u)$
is an even function of $u$, there is another root at $u=-2$. Tracing back the transformations,
we find the amplitude mode frequencies at $$E=i\omega=i2|\Delta_{{\bf Q}}|z=-2|\Delta_{{\bf Q}}|u
=\pm4|\Delta_{{\bf Q}}|=\pm2|\Delta_{{\rm DDW}}|.$$  The last equality follows by 
recalling the DDW single-particle gap at $(\pi,0)$, 
$\Delta_{{\rm DDW}}=2|\Delta_{{\bf Q}}|$. So, we conclude that the amplitude modes for the DDW state lie
at frequencies $\pm 2\Delta_{{\rm DDW}}$. 

        At the level of our computation, the mode is undamped, i.e., an exact eigenstate
of the system. But, since the ordered state is a particle-hole condensate in contrast
to a superconductor, it can easily couple to
other degrees of freedom in the cuprates. The mode can, for example, couple to the optical 
phonons, which are abundant in the system. 
These other degrees of freedom will act as an external bath to the {\it system} of the amplitude 
boson, and provide it with some frequency width.  
We can meaningfully talk about the mode - and
its manifestations in experiments - only as long as this width is not too big.

  
 In order to make contact with the neutron scattering experiments, let us start with 
  the interaction potential between the incoming neutron and the electrons carrying
  the orbital current, $V_{int}$,\cite{Hsu} 
  
 \begin{equation}
V_{int} = \sum_{\langle i j \rangle} t c_{i \sigma}^{\dagger} c_{j \sigma}
\exp{(ie \int_i^j {\bf A} \cdot d{\bf l} ) }+{\rm h.c.}
\end{equation} 
 
 Here, ${\bf A}$ is the electromagnetic gauge potential generated by the neutron
 magnetic moment ${\bf \mu_{n}}$,
 ${\bf A} = \mu_{n} \times ({\bf r}_e-{\bf r}_n)/(|{\bf r}_e-{\bf r}_n|^3),$ 
  $\mu_{n}= -1.91 \frac{e }{m_N  } {\bf S}$, and,
${\bf r}_e$ and ${\bf r}_n$ are the electron and neutron coordinates, respectively.
   Because the gauge field generated by the neutron is weak, we can approximate $V_{int}$ by,
  $V_{int} =iet \sum_{\langle i j \rangle} 
( \int_i^j {\bf A} \cdot d{\bf l} )( c_{i \sigma}^{\dagger} 
c_{j \sigma}-c_{j \sigma}^{\dagger} c_{i \sigma}).$

 To find the cross section, we first have to evaluate the matrix element of the interaction potential
 between the initial and final states of the neutron,
 $\langle{\bf k}_f|V_{int}|{\bf k}_i\rangle=\int d^3 
{\bf r}_nV_{int}\exp{i{\bf q}\cdot{\bf r}_n}$,
where, $ {\bf q} = {\bf k}_i-{\bf k}_f$. 
 After doing the integral over the neutron coordinates,\cite{Hsu} summing over the initial and final
 states in the scattering process, and then, putting ${\bf q}={\bf Q}=(\pi,\pi)$, we find, 
 \begin{equation}
 \langle{\bf k}_f|V_{int}|{\bf k}_i\rangle = \frac{8ite}
 {\pi}\mu_{z}\sum_{{\bf k},\sigma}(\cos k_x-\cos k_y)
 c_{{\bf k}+{\bf Q},\sigma}^{\dagger}c_{{\bf k},\sigma}.
 \label{mat}
 \end{equation} 
 
 Apart from some coefficients, the expression on the right hand side of the last equation is
 precisely the DDW gap in the operator form. In other words, the coupling of the neutron with
 the system at $(\pi,\pi)$ is, as expected, through the orbital currents. 
 
 Finally, averaging over the neutron spin states 
 using $\sum_{\sigma_i \sigma_f} p_{\sigma_i} 
  \langle \sigma_i | \mu_{\alpha}
  | \sigma_f \rangle 
  \langle \sigma_f | \mu_{\beta}
  | \sigma_i \rangle = \mu^2 \delta_{\alpha \beta},$ and then,
  follwing standard steps,\cite{Lovesay} we derive, 
 
 \begin{equation}
\frac{d^2\sigma}{d\Omega dE_f}
=(\frac{8\mu te}{\pi g})^{2}\langle \widehat{\Delta}_{{\bf Q},\omega}\widehat{\Delta}_{-{\bf Q},-\omega}\rangle.
\label{xsection}
\end{equation}
Here, $\widehat{\Delta}_{{\bf Q},\omega}$ is the frequency transform of the gap parameter, $\widehat{\Delta}_{{\bf Q}}$,
  given, in operator form for a general ${\bf q}$, in Eq.~\ref{Delta}.
 
 The expression on the right hand side of the last equation is precisely the quantity we
 evaluated in the analysis of the DDW amplitude mode. The quantity
 $ \langle \widehat{\Delta}_{{\bf Q},\omega}\widehat{\Delta}_{-{\bf Q},-\omega}\rangle$, in imaginary time ordered form, 
 can be directly taken over from Eq.~\ref{Perturb action2},
  \begin {equation}
  \langle T_{\tau}\widehat{\Delta}_{{\bf Q},\omega}\widehat{\Delta}_{-{\bf Q},-\omega}\rangle = \frac{1}
  {A_{{\bf Q},\omega}+A_{-{\bf Q},-\omega}+
2\widetilde{B}_{{\bf Q},\omega}}
\end {equation}
 
 As we found earlier, this function has poles at 
 $\omega=\pm i2\Delta_{{\rm DDW}}$. These are the dissipative
 and the absorptive responses of the system, respectively.
 At zero temperature, the relevant pole is that for the dissipative response only. 
 After analytically continuing the frequency, the corresponding retarded function will have
 a pole at $\omega= 2\Delta_{{\rm DDW}}$,
 implying a sharp peak at that frequency. This peak
captures the amplitude boson for the DDW state in inelastic neutron scattering. As we emphasized before,
the sharpness of the response is really an artifact of our calculation, which excludes
the coupling of the boson to other degrees of freedom, such as the optic phonons. A fuller treatment should bring in some 
damping to the mode. Indeed, the peaks above $T_{c}$ in the neutron scattering of the cuprates 
are broad.\cite{FongMook} Clearly, more work is needed in this direction 
to establish this. To establish the correct momentum dispersion will require a more microscopic 
model than that considered. From translational invariance one can argue that the square of the
mode frequency is an analytic function of the momentum, that is, it must begin quadratically. However, the coefficient of
the leading  momentum dependence is expected\cite{popov} to be complex implying damping of the collective mode at a finite
wavevector away from ${\bf Q}=(\pi,\pi)$.
\acknowledgements
This work was supported by a grant from the National Science Foundation: DMR-9971138. We thank C. Nayak for interesting
discussions. One of us (S. C.) would also like to thank L. Yin for discussions on a similar problem of the Higgs
mode in superconductors.\cite{Lan}

\end{document}